\documentstyle[preprint,aps,psfig]{revtex}
\begin{document}
\draft
\preprint{}
\title{Distillation of Strangelets for low initial $\mu/T$}
\author{
C.~Spieles$^a$, C.~Greiner$^b$, H.~St\"ocker$^a$, and J.~P.~Coffin$^c$}
\address{ $^a$ Institut f\"ur
Theoretische Physik,  J.~W.~Goethe-Universit\"at,
60054 Frankfurt
am Main, Germany}
\address{$^b$ Department of Physics, Duke University, Durham, NC 27708,
U.S.A}
\address{$^c$ Centre de Recherches Nucl\'{e}aires de Strasbourg
et Universit\'{e} Louis Pasteur Strasbourg, France}
\date{\today}
\maketitle
\begin{abstract}
We calculate the evolution of quark-gluon-plasma droplets during the
hadronization in a thermodynamical model. It is
speculated that cooling as well as strangeness enrichment allow for
the formation of
strangelets even at very high initial entropy per baryon $S/A^{\rm
init}\approx 500$ and low initial
baryon numbers of $A_{\rm B}^{\rm init}\approx 30$. It is shown that
the droplet  with vanishing initial
chemical potential of strange quarks and a very moderate chemical potential
of up/down quarks immediately charges up with
strangeness. Baryon densities of $\approx 2\rho_0$ and
strange chemical potentials of $\mu_s>350$~MeV are reached
if strangelets are stable.
The importance of
net--baryon and net--strangeness fluctuations for the possible strangelet
formation at RHIC and LHC is emphasized.
\end{abstract}

\narrowtext

Strangelets can be thought of as
strange multiquark clusters which should
be more compressed than ordinary nuclei and may
exist as (meta-)stable exotic isomers of nuclear matter \cite{Bod71}.
It was speculated \cite{Wi84} that strange matter might
resolve the dark matter issue, if it would be absolutely stable.

The possible creation --- in heavy ion collisions ---
 of such long-lived remnants of the quark-gluon-plasma, cooled and charged
up with strangeness by the emission of pions and kaons, was proposed by Liu
and Shaw \cite{Liu84} and Greiner {\it et al.} \cite{CG1,CG2}.
Thus, strangelets can serve as unambiguous signatures for the
creation of a quark gluon plasma.
The detection of strangelets would verify exciting
theoretical ideas with consequenses for our knowledge of
the evolution of the early universe, the dynamics of supernova explosions
and the underlying theory of strong interactions \cite{P91}.

Here we want to point out that
such exotic states of matter can be created in
heavy ion collisions even at collider energies, where such a process has
received no attention so far, because common belief was that the (strange)
baryon densities vanish at midrapidity, both at RHIC and LHC. We argue,
however, that this conclusion was premature. This is due to the following
effects:
\begin{itemize}
\item fluctuations of the stopping power can provide finite
      baryochemical potential $\mu_{\rm B}$ at mid--rapidity
      in a small fraction of all events;
\item fluctuations of the net--baryon and --strangeness content between
      different rapidity bins within {\em one} event can be quite large;
\item strange (anti--)baryon enhancement due to collective effects
      (e.~g. a chiral phase transition);
\item strangeness and baryon distillery, which are inherent for
    the two--phase system (hadron gas/quark gluon plasma) for a
    wide parameter range.
\end{itemize}
The last point stresses the significance of the 'chemistry` of the system
in the evolution of the phase transition.

In the following we adopt a model \cite{CG2} for the dynamical creation of
strangelets
via the strangeness separation mechanism\cite{CG1}.
Consider a first order phase transition of the QGP to hadron gas.
Strange and antistrange quarks do not hadronize
at the same time for a baryon--rich system\cite{CG1}.
The separation mechanism can be viewed as being due to the associated
production of kaons (containing $\bar s$ quarks) in the hadron phase,
because of the surplus of massless quarks compared to their
antiquarks. The strange quarks can combine
to $\Lambda $-particles, but it is
energetically favourable that s-quarks remain in the plasma, when
hadronization proceeds.
The ratio $f_s$
 of the net strangess over the net baryon number
 quantifies  the
excess of net strangeness.
Both the hadronic and the quark matter phases enter the strange sector
$f_s\neq 0$ of the phase
diagram almost immediately, which
has up  to now been neglected in almost all calculations of the time
evolution of the system.
Earlier studies addressed the case of a baryon--rich QGP
with rather moderate entropy per baryon \cite{CG1,CG2,Lee93}.
Now we focus on {\em low
initial baryon densities} and {\em high specific entropies}, to match
the expected conditions of heavy ion collisions at RHIC and LHC, where the
search for strangeness enters in the objective of the ALICE experiment
\cite{ALICE}.

The hadronization transition has been
described by geometric and statistical models,
where the matter is assumed to be in partial or complete equilibrium
during the whole \mbox{(quasi-)}isentropic expansion.
A more realistic scenario must take
into account the particle radiation from the surface of the hadronic
fireball before `freeze out'.
Our model \cite{CG2} combines
these two pictures. The expansion of the QGP
droplet during the phase transition is described as a  two-phase
equilibrium; in particular the
strangeness degree of freedom stays in chemical equilibrium
because the complete hadronic particle production is driven
by the plasma phase.
The nonequilibrium radiation is incorporated by rapid
freeze-out of hadrons from the outer layer of the hadron phase
surrounding the QGP droplet.
During the expansion, the volume increase of the system thus
competes with the decrease due to the freeze--out.
The global properties like (decreasing) $S/A$ and (increasing) $f_s$
 of the remaining two-phase system then change in time
according to the following differential equations for the baryon number, the
entropy, and the net strangeness number of the total system:
\begin{eqnarray}\label{eq1}
\frac{d}{dt}A^{tot}  & = & -\Gamma \, A^{HG}  \nonumber \\
\frac{d}{dt}S^{tot}  & = & -\Gamma \, S^{HG} \\
\frac{d}{dt}(N_s - N_{\overline{s}})^{tot}  & =  & -\Gamma \,
(N_s - N_{\overline{s}})^{HG} \, \, \, , \nonumber
\end{eqnarray}
where $\Gamma = \frac{1}{A^{HG}}
\left( \frac{\Delta A^{HG}}{\Delta t} \right) _{ev}$ is the effective
(`universal') rate of particles (of converted hadron gas volume)
evaporated from the hadron phase.
The equation of state consists of the bag model for the
quark gluon plasma and a
mixture of relativistic Bose--Einstein and Fermi--Dirac gases of well
established strange and non--strange hadrons up to
in Hagedorn's eigenvolume correction for the
hadron matter \cite{CG1}.
Thus, one solves simultaneously
the equations of motion (\ref{eq1}) and the Gibbs phase equilibrium
conditions for the intrinsic variables, i.e. the chemical potentials and the
temperature, as functions of time.

\begin{figure}
\centerline{\psfig{figure=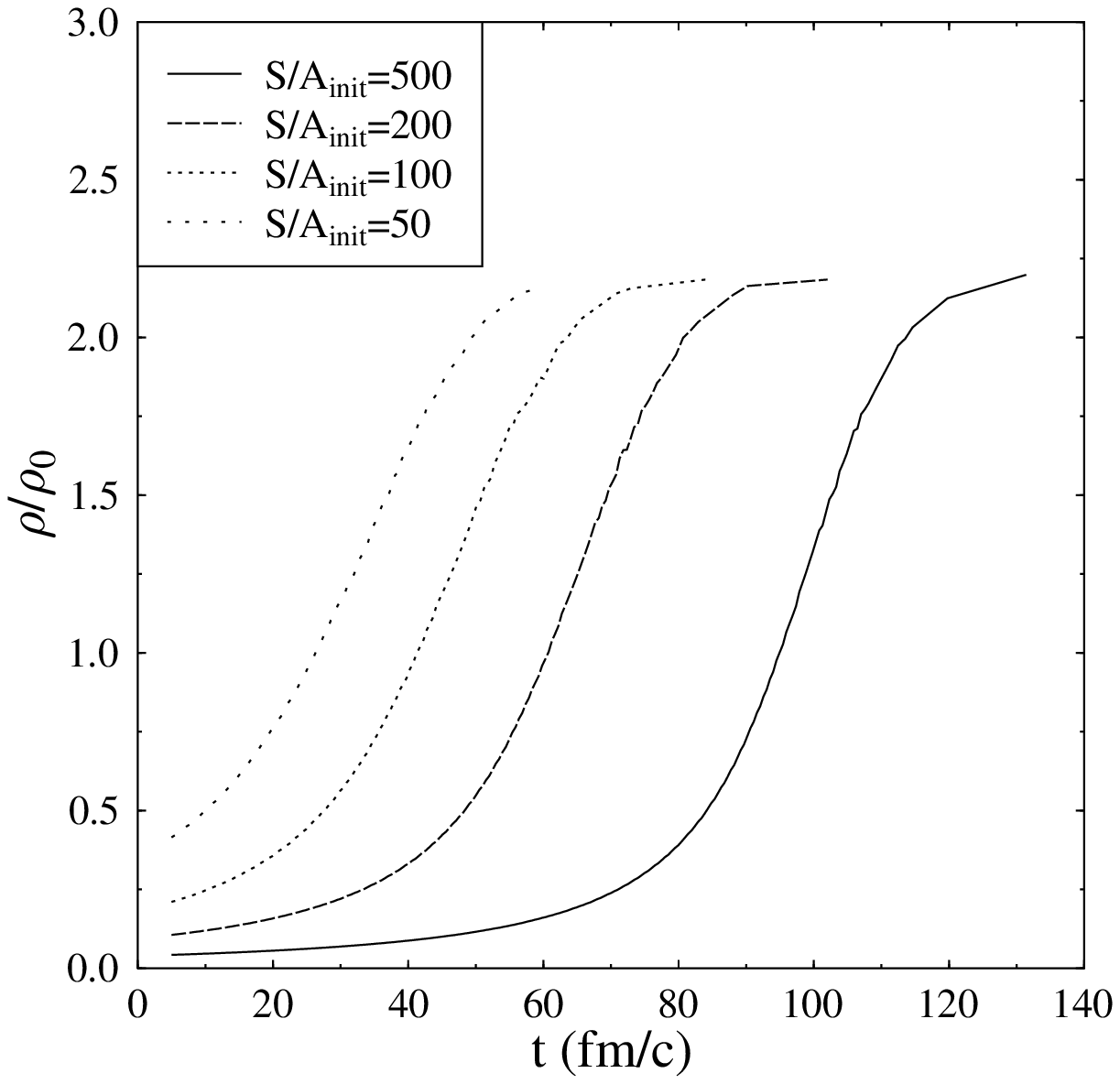,width=13cm}}
\caption{Time evolution of the net baryon density of a QGP droplet.
The initial conditions are
$f_s^{\rm init}=0$ and $A_{\rm B}^{\rm
init}=30$. The bag constant is $B^{1/4}=160$~MeV.
\label{tdens}}
\end{figure}

Fig.~\ref{tdens} illustrates
the increase of {\em baryon concentration} in the plasma droplet
as an inherent feature of the dynamics of the phase
transition (cf. \cite{Wi84}).
The origin of this result lies in
 the fact that the baryon number in the
quark--gluon phase is carried by quarks with $m_{\rm q}\ll T_{\rm C}$, while
the baryon density in the hadron phase is suppressed by a Boltzmann factor
$\exp (-m_{\rm baryon}/T_{\rm C})$ with $m_{\rm baryon}\gg T_{\rm C}$.
Mainly mesons (pions and kaons) are created
in the hadronic phase.
More relative entropy $S/A$ than baryon number is carried away in the
hadronization and
evaporation process\cite{CG2}, i.e.
$(S/A)^{HG} \gg (S/A)^{QGP}$. Ultimately, whether $(S/A)^{HG}$ is larger or
smaller than $(S/A)^{QGP}$ at finite, nonvanishing chemical potentials might
theoretically only be proven rigorously by lattice gauge calculations in the
furure. However, model equations of state do suggest such a behaviour,
which would open such intriguing possibilities as baryon inhomogenities in
ultrarelativistic heavy ion collisions.
In the early universe
shrinking quark droplets may --- in analogy ---
contain the accumulated baryon number
with possibly very high baryon density \cite{Wi84}. This
mechanism yields a primeval inhomegeneous
nucleosynthesis \cite{P91,App85}, which is signaled by the abundances of the
light elements.

What `initial' conditions do we expect at collider energies?
At RHIC energies one might see
baryon stopping,
$dN_B/dy >0$, on the average, at midrapidity. This can be
due to multiple rescattering,
 leading to
a nonvanishing, positive quarkchemical potential $\mu_q$ \cite{Du93}.
On the other hand, relativistic
meson--field models, which, at high temperature,
qualitatitively simulate chiral behaviour of
the nuclear matter, exhibit a transition into a
phase of massless baryons \cite{The83}.
Including hyperons and $YY$--interaction \cite{Sch92}
it shows that at $\mu\approx 0$
the densities for all baryon species are of the order of $\rho_0$ near
the critical temperature.
Thus, the fraction of \mbox{(anti--)}strange quarks
increases drastically.
Several hundred \mbox{(anti--)}baryons, many of them
\mbox{(anti--)}hyperons  may then fill the hot
 midrapidity  region
(with net baryon number $\approx 0$).
Districts of non--vanishing net baryon (respectively
anti--baryon) density with finite $s$ ($\bar s$) content will then occur
stochastically. Thus, the finite chemical potential is locally caused by the
fluctuations of newly produced particles, not by the stopped matter.
If such a phenomenon persists also for the deconfined phase,
the effect of baryon concentration
and strangeness separation
may then result in the production of strangelets and anti-strangelets
in roughly equal amounts.

Let us try to give a rough estimate of the possible size of
fluctuations in the net baryon and net strangeness
number to be expected at midrapidity
(or fluctuations along different rapidity intervals) around their mean.
The average number of initial quarks and antiquarks
(before hadronization)
in a rapidity interval is approximately $1/2 dN_{\pi}/dy \cdot \Delta y$, if
half of the pions are made by the quarks and the other half by the gluons.
For RHIC energies $dN_{\pi}/dy$ has been estimated to lie between 1100 and
1600
and for LHC energies to lie up to nearly 4000 in central Au+Au collisions
\cite{Wa91}.
Hence the quark number is roughly 500 for RHIC and up to 2000 for LHC
in a rapidity interval $\Delta y \sim 1$.
(In an equilibrated plasma the total number of quarks is
$N\sim \rho_q \pi R^2 \Delta z$
within $\Delta z=1-2$~fm in the early stage of the hydrodynamical expansion.
According to
$\rho _q = g \frac{3}{4\pi^2}T^3\zeta(3)\approx 1.1 T^3 $
for a degeneracy of g=12 these numbers correspond to temperatures of
$ T\sim 250 - 500$ MeV.)
We take the number to be $N=500$.
A similar consideration holds for
strange and antistrange quarks, and we take here $N_s=200$.
We now assume independent fluctuations according to Poissonians
within this rapidity interval.
In fact the actual width of the fluctuation
at collider energies could be much broader (KNO--scaling of particle
multiplicity distributions in elementary $pp$--collisions\cite{UA5}).
To justify the assumption of independent
fluctuations of $B$ and $\bar B$ despite of the local compensation of quantum
numbers, one has to estimate the typical relative momenta within a
quark--antiquark pair.
If one follows the parton
cascade concept embodying perturbative QCD\cite{Geig}, the average
$\sqrt{\hat s}$ of first parton--parton interactions ($gg\rightarrow q\bar
q$ being the most important contribution) should be of the order
of $5-10$~GeV at LHC energies.
The produced $B$ and $\bar B$, carrying about $0.4$ of the
\mbox{(anti--)}quark momenta, would thus be separated in rapidity by at least
 one unit (assuming a transverse momentum of about 500~MeV).

The net baryon number
in the box described above will be $|B|>30$ with a probability of 0.5~\%.
About $0.1$~\% of the events will reach $|B|>30$ with a strangeness fraction
$|f_{\rm s}|>0.7$. Hence fluctuations are not negligible.
If each pion carries about 3.6 units of entropy
(which is true for massless bosons), the entropy per baryon content
in the fireball is
\begin{equation}
\frac{S}{A_B} \, \approx \, 3.6 \, \frac{dN_{\pi}/dy}{dN_{B}/dy}
\,\,\, ,
\end{equation}
and thus for $dN_B/dy = 30$ a range of 60 to 250 is formed.
 If the plasma is equilibrated, the ratio of the
quarkchemical potential and the temperature $|\mu | /T$ is directly related
to the entropy per baryon number via
\begin{equation}
\left(\frac{S}{|A_B|}\right)^{QGP} \, \approx \, \frac{37}{15} \pi^2 \,
\left(\frac{|\mu |}{T}\right)^{-1}
\,\,\, .
\end{equation}
Accordingly the ratio then varies between 0.1 to 0.4.

 We now consider various
fireballs with an initial net baryon number of $A_B=30$ and
a net strangeness fraction $f_s$ of either 0 or 0.7.
The initial entropy per baryon ratios are chosen between 50 and 500.
Table 1 summarizes the initial conditions
(adjusted by the (initial)
chemical potentials $\mu_q, \,\mu_s$ and temperature) used to start
the hadronization.
It also shows the final parameters of the quark droplet
like the saturated strangeness content and baryon number.
One further, yet crucial input, is the bag constant employed to describe
the equation of state of the (strange) quark matter droplet.
Only for the bag constants $B^{1/4} \leq 180 $ MeV strange matter does
exist as a metastable state at zero temperature \cite{CG1}, being absolutely
stable only  for $B^{1/4} < 150 $ MeV \cite{Wi84}.

\begin{figure}
\centerline{\psfig{figure=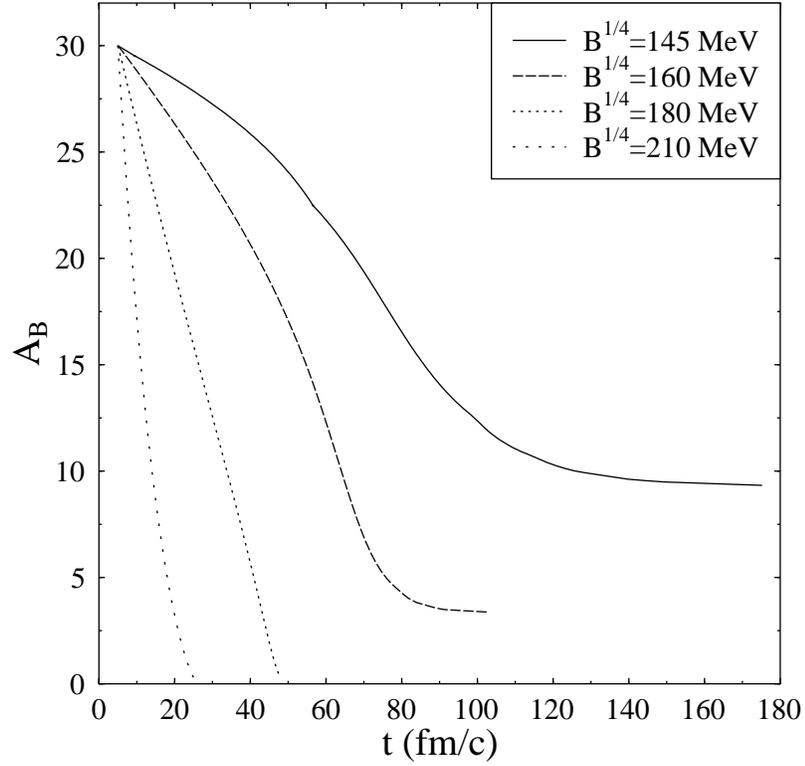,width=13cm}}
\caption{Time evolution of the baryon number for
a QGP droplet with $A_{\rm B}^{\rm init}=30$, $S/A^{\rm init}=200$,
$f_s^{\rm init}=0.7$
and different bag constants.
\label{ab_b14}}
\end{figure}

Fig.~\ref{ab_b14} shows the time evolution of the baryon number for
 $S/A^{\rm init}=200$ and $f_s^{\rm init}=0.7$ for  varyous
bag constants.
For $B^{1/4}<180$~MeV
a cold
strangelet emerges from the expansion and evaporation process, while
the droplet completely hadronizes for bag constants $B^{1/4}\ge 180$~MeV
(for $B^{1/4}=210 $ MeV hadronization proceeds
without any significant
cooling of the quark phase, although the specific entropy $S/A$
decreases by a factor of 2 from 200 to only 100).
The strangeness
separation works also in these cases, as can be read off the
large final values of the
net strangeness content,
$f_s \stackrel{>}{\sim } 1.5-2$. However, then the volume of the drop
becomes small, it decays and the
strange quarks hadronize
into $\Lambda $-particles and other strange hadrons.

\begin{figure}
\centerline{\psfig{figure=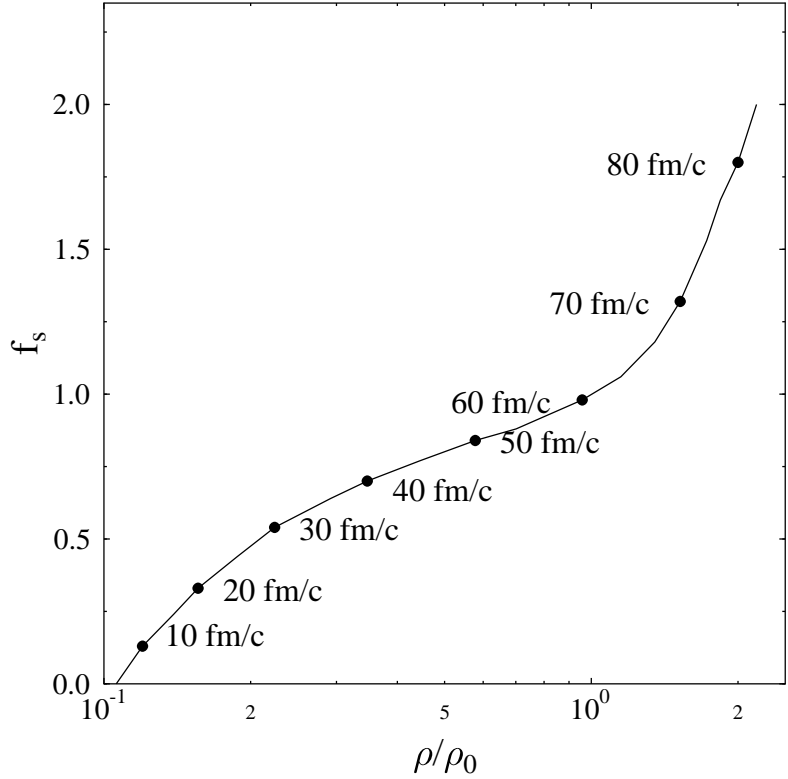,width=13cm}}
\caption{Evolution of a QGP droplet with baryon number $A_{\rm B}^{\rm
init}=30$ for $S/A^{\rm init}=200$ and
$f_s^{\rm init}=0$. The bag constant is $B^{1/4}=160$~MeV. Shown is the
baryon density and the corresponding strangeness fraction.
\label{fsrho}}
\end{figure}

Fig.~\ref{fsrho} shows the evolution of the two-phase system for $S/A^{\rm
init}=200$, $f_s^{\rm init}=0$ and for a bag constant $B^{1/4}=160$~MeV
in the plane of the strangeness fraction vs. the baryon density.
The baryon density increases by more than one order of magnitude!
Correspondingly, the chemical potential rises as drastically
during the evolution, namely from $\mu^i=16$~MeV to $\mu^f>200$~MeV.
The strangeness
separation mechanism drives the chemical potential of the strange quarks
from $\mu^i_s=0$ up to $\mu^f_s\approx 400$~MeV.
Thus, the thermodynamical and
chemical properties during the time evolution
are quite different from
 the initial conditions of the system.

Even for high initial entropies, $S/A\approx100-500$,
in the quark blob
the entropy in the remaining droplet
approaches zero at the end of the evolution
(assuming $B^{1/4}=160$~MeV).
High initial entropies per baryon
require more time for kaon and pion evaporation in order to end up finally
with the same configuration of (meta--)stable strange quark matter.

In conclusion, we have shown in the present model that
the evolution of quark-gluon-plasma droplets during their
hadronization may result in the formation of
strangelets even at very high initial entropy per baryon $S/A^{\rm
init}\approx 500$ and low initial
baryon numbers of $A_{\rm B}^{\rm init}\approx 30$.
The distillation
of very small strangelets of a size $A_B \leq 10$ (see Table 1) is possible.
We note
that finite size effects of describing small strangelets
neglected here might become
of crucial importance \cite{Mad93}.
Special (meta-)stable candidates are the quark-alpha \cite{Mi88} with $A_B=6$
and the H-Dibaryon state with $A_B=2$ \cite{Jaf77}.
Local net--baryon and net--strangeness fluctuations can provide
suitable initial conditions for the possible strangelet creation at
RHIC and LHC.
Droplets  with  vanishing initial
chemical potential of strange quarks and a small chemical potential
of up/down quarks quickly charge up with
strangeness and baryon--number: if strangelets are stable,
the droplet reaches strange chemical potentials of $\mu_s>350$~MeV and
two times ground state nuclear matter density!

\acknowledgements

C.G. thanks the Alexander von Humboldt Stiftung
for his support by a Feodor Lynen scholarship.
This work was supported
in part by the U.S. Department of Energy (grant DE-FG05-90ER40592), by the
Gesellschaft f\"ur Schwerionenforschung, Darmstadt, Germany,
the Bundesministerium
f\"ur Forschung und Technologie (F.~R.~G.), Deutsche Forschungsgemeinschaft
and the Graduiertenkolleg Schwerionenphysik.

%

%--------------------------------------------------------------------
%--------------------------------------------------------------------
%--------------------------------------------------------------------
%--------------------------------------------------------------------

\widetext

\begin{table}
\caption{Various situations of an hadronizing plasma droplet.
In the first column the bag constant for describing the plasma phase
is listed. Then the initial conditions follow. The final values
for the baryon number, the strangeness fraction and the two chemical
potentials
at the end (or after) hadronization are listed in the last four columns.}

\begin{tabular}{|c||ccccccc|ccccc|}\hline
$B^{1/4}  $ & $S/A^{\rm init}$ & $f_s^{\rm init}$ & $A_{\rm B}^{\rm init}$ &
$\rho_B^{\rm init}$  & $\mu_q^{\rm init}$ & $\mu_s^{\rm init}$
& $\mu/T^{\rm init}$ &  $f_s^{\rm final}$
& $A_{\rm B}^{\rm final}$ & $\rho_B^{\rm final}$
&$\mu_q^{\rm final}$ & $\mu_s^{\rm init}$ \\
$ ({\rm MeV}) $ &  &  &  &
$ ({\rm fm^{-3}})$ & ({\rm MeV}) & ({\rm MeV})
&  &
&
&$ ({\rm fm^{-3}})$
&({\rm MeV})  & ({\rm MeV})
\\ \hline \hline
160  & 500 & 0   & 30 &0.0067& 6.55&0 & 0.060  & 1.99  & 2.36
&0.352 &224.3&396.0\\
160  & 500 & 0.7 & 30 &0.0068& 5.02&4.01 & 0.046  & 1.96&2.71
&0.339&216.9&386.0\\
160  & 200 & 0   & 30 &0.017 & 16.36&0 &0.150      & 2.0 &2.79
&0.349&224.7&396.9\\
160  & 200 & 0.7 & 30 &0.017 &12.55&10.04&0.115&2.0&3.37&0.350
&223.7&396.3\\
160  & 100 & 0   & 30 &0.034 &32.60 & 0&0.300&1.99&2.93&0.350
&225.2&396.7\\
160  & 100 & 0.7 & 30 &0.034 &25.07&20.09&0.185&2.0&3.96&0.352
&223.4&396.2\\
160  & 50  & 0   & 30 &0.066 &64.23&0&0.599&1.94&2.75&0.344
&219.7&385.8\\
160  & 50  & 0.7 & 30 &0.067&49.81&40.26&0.463&1.99&4.56&0.350
&223.3&395.6\\
145  & 200 & 0.7 & 30 &0.012 &11.23&9.52&0.114&1.60&9.33&0.270
&234.6&347.6\\
180  & 200 & 0.7 & 30 &0.024&14.20&10.76&0.116&(1.83)&0&(0.349)
&(146.8)&(315.7)\\
210  & 200 & 0.7 & 30 &0.039&16.74&11.99&0.117&(1.58)&0&(0.063)
&(19.5)&(50.5)\\
\hline
\end{tabular}
\end{table}
\end{document}